 \pdfoutput=1 
 \documentclass[twoside]{IEEEtran}

 \IEEEoverridecommandlockouts
 \usepackage{algorithm}

\usepackage{tabularx}
\UseRawInputEncoding

\usepackage{array}
\usepackage{mdwmath}
\usepackage{mathtools}
\usepackage{subfig}
\usepackage{tabularx}
\usepackage{graphicx}
\usepackage{url}
\usepackage[noend]{algpseudocode}
\graphicspath{{../pdf/}{../jpeg/}}
\DeclareGraphicsExtensions{.pdf,.jpeg,.png}

\usepackage{amsfonts}
\usepackage{booktabs}
\usepackage{lipsum}

\usepackage{tikz}
\usetikzlibrary{%
patterns,%
calc,%
fit,%
arrows,%
plotmarks,%
shadows,%
chains,%
shapes%
}
\usepackage{pgfplots}
\usepackage{cite}
\usepackage{amsmath,amssymb,amsfonts}

\newtheorem{remark}{Remark}

\usepackage{latexsym}
\usepackage{amsfonts,amssymb,amsmath}
\usepackage[normalem]{ulem}

\title{Delay-Augmented Stacked Intelligent Surfaces: Potential, Challenges, and Opportunities}

\author{
\IEEEauthorblockN{Hibatallah Alwazani, \IEEEmembership{Student Member, IEEE}, Omran Abbas, \IEEEmembership{Student Member, IEEE}, Lo\"ic Markley, \IEEEmembership{Senior Member, IEEE}, Anas Chaaban, \IEEEmembership{Senior Member, IEEE}}\\

\thanks{%
The authors are with the School of Engineering, the University of British Columbia, 1137 Alumni Ave., Kelowna, BC V1V1V7, Canada (email: anas.chaaban@ubc.ca)}
}
\begin{document}

\maketitle

\begin{abstract}
    Stacked intelligent surfaces (SIS)s have been proposed recently as {an enabling technology for}  Holographic {Multiple} Input {Multiple} Output (HMIMO) and Ultra-massive MIMO (umMIMO) technologies. Their utility can extend beyond spatial wave-domain processing of signals if they are enhanced with strategically-tuned  {symbol-duration level} delays to enable temporal processing as well. In this work, we {introduce the idea of} a delay-augmented SIS (DA-SIS). We shed light on the feasibility of realizing delay units in an SIS. Then, we discuss the relevance of the proposed DA-SIS and present a use case that illustrates its potential, wherein the DA-SIS serves as an analog equalizer that aids in eliminating  {multi-path-induced} inter-symbol-interference (ISI). We show how the number of elements affect the equalization process using the bit error rate (BER) as a metric, and demonstrate the potential of the DA-SIS in equalization via comparing with  digital equalizers as a benchmark. 
Finally, we present opportunities and future research directions that can be undertaken to bring this idea to fruition. 
\end{abstract}

\section{Introduction}

Building on the success of massive MIMO, ultra-massive MIMO (umMIMO) is envisioned as a foundational wireless technology for beyond-5G systems, addressing the ever-growing demand for higher data rates and reliability through using dense antenna configurations. The asymptotic limit of umMIMO (in terms of antenna density) is well approximated by Holographic MIMO (HMIMO), which has recently emerged as a promising paradigm. HMIMO features a spatially continuous, antenna-dense aperture that offers enhanced beamforming gains, reduced interference, and improved spatial multiplexing capabilities \cite{emilHMIMO}.

A practical and cost-effective implementation of HMIMO is  {envisioned} through  {the use of } Stacked Intelligent Surfaces (SISs) 
 {also called Stacked Intelligent Metasurfaces (SIMs)}\cite{6gsim}. SISs consist of multiple layers of reconfigurable {surfaces} capable of manipulating electromagnetic (EM) waves in space. By offloading some transceiver tasks from the electronic to the EM domain, SISs reduce the complexity and power consumption traditionally associated with digital transceivers. This makes SIS-based architectures particularly attractive for  {enhancing scalability and energy-efficiency.}

SISs can be viewed as an evolution of Reconfigurable Intelligent Surfaces (RIS), which are composed of passive elements that {can be optimized} to enhance communication performance \cite{annieSIM}. While RIS  has demonstrated potential, it faces several practical limitations, such as high overhead in channel estimation, sensitivity to placement due to double path loss, and increased network complexity due to the inclusion of another node in the network \cite{annieSIM}. 

By contrast, SISs, when integrated directly onto the transceiver, circumvent many of these limitations. This integration allows for simplified medium access control, more efficient beamforming, and simplified transceiver design. In this context, SISs are emerging as a superior alternative to RIS, paving the way for more efficient and practical implementations of umMIMO and HMIMO systems.

 Next, we outline several notable works using the SIS.
\subsection{State of the Art}
{Capitalizing on their wave-domain beamforming capabilities, SISs have been investigated in the literature for communication and sensing applications. This feature was utilized in \cite{SIMbeam} to maximize the received signal power at a desired location, and to create {interference-free} spatial sub-channels in \cite{6gsim} via diagonalizing the channel to the largest extent possible using an SIS. It was also utilized in \cite{annieSIM} to aid in channel estimation in an SIS-assisted  {Multiple Input Single Output (MISO)} system \cite{annieSIM}. 
There are several other works that have investigated the utility of SISs in communications and sensing, and the readers are referred to \cite{liu2024stackedintelligentmetasurfaceswireless} for an overview of its capabilities and limitations.} 

All above and current works interpret the SIS as an architecture that implements spatial processing (precoding/preprocessing at the transmitter and/or combining/postprocessing) at the receiver. This motivates the following question: Is the SIS limited to only performing spatial processing? Discussing this question is the main focus of this paper.

\subsection{Contribution}
What if the SIS could affect  not just the spatial characteristics of an incoming signal, but also its temporal characteristics? In this work, we introduce an architectural enhancement by incorporating a temporal dimension into the SIS through the integration of time-delays  directly onto the SIS architecture.  While conventional SISs manipulate the wavefront through phase shifts corresponding to delays comparable to fractions of the carrier period, we propose a delay-augmented SIS (DA-SIS) that extends this capability by introducing programmable delays on the order of a symbol duration, thus enabling dynamic temporal control over signal propagation \footnote{{The two types of delays are categorized separately to reflect their markedly different levels of complexity and practical feasibility. }}.


\begin{figure*}
    \centering
\includegraphics[width=\linewidth]{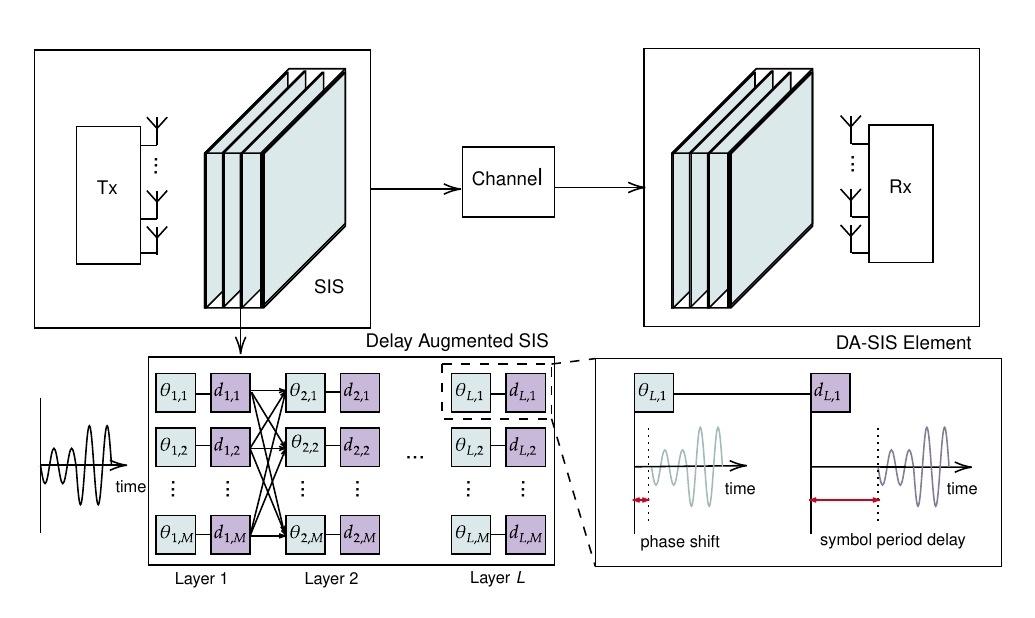}
    \caption{SIS-assisted communication system and SIS architecture. The blocks denote phase shifts ($\theta_{i,j}$, introducing sub-wavelength-level phase shifts) and time delay ($d_{i,j}$ introducing symbol-duration-level delays). For a conventional SIS, $d_{i,j}=0$. }
    \label{Delayfig}
\end{figure*}

This advancement lays the foundation for a new class of reconfigurable surfaces capable of spatiotemporal signal processing. Our key contributions are as follows:
\begin{enumerate}
    \item We conceptualize a Delay-Augmented SIS (DA-SIS) architecture that integrates temporal processing capabilities into the conventional SIS.
    \item  We shed light on the feasibility, operational relevance, and performance benefits of the proposed DA-SIS.
    \item We illustrate the potential of the DA-SIS in a practical scenario, showcasing its ability to perform  equalization of signals with multi-path-induced inter-symbol interference (ISI).
\item We discuss emerging challenges posed by this architecture and outline promising directions for future research.
\end{enumerate}

Next, we delve into the architectural design, operational principle, and potential physical realization of the proposed DA-SIS.

\section{Delay-Augmented SIS}

 We begin by outlining the architecture and overall operation of the DA-SIS, followed by an analysis of its physical feasibility and a discussion of various implementation strategies of the delay feature.

Fig. \ref{Delayfig} illustrates an SIS-assisted communication system. In conventional SIS architectures, the {elements in each layer} impose phase shifts ($\theta_{i,j}$ in Fig. \ref{Delayfig}; layer $i$, element $j$) on incident EM waves, thereby enabling spatial wavefront manipulation. In contrast, the proposed DA-SIS architecture extends this capability by superimposing a controllable time delay ($d_{i,j}$ in Fig. \ref{Delayfig}; layer $i$, element $j$) onto the phase shift at each element.
This introduces a new degree of freedom in signal shaping and propagation control. In the following subsections, we examine the design considerations for integrating time-delay elements into the {structure of each SIS layer}, evaluate their practical feasibility, and overview methods for the hardware realization of these delay units.
\vspace{-0.1in}
\subsection{DA-SIS Architecture}

The main operation principle of an SIS is to introduce time delays to the EM wave propagating through it. This is widely used in electromagnetism and optics to realize different objectives such as beam-forming and modulation. We can divide the delay operation into short delays (carrier period level) and long delays (symbol duration level). 

\subsubsection{Short time delay} A short delay is widely used to ensure EM waves interfere constructively or destructively at specific points or regions in space. Typically, a short delay is smaller than the time required for an EM wave to travel a distance equal to its wavelength. For example, in antenna arrays, a very short time delay (typically in the range of picoseconds in millimetre {wave} communication systems) is applied to generate different beam patterns. Such delays can be realized using circuit phase shifters (transmission lines with different lengths) or {antenna arrays} \cite{TRaArray}. In an array, elements provide phase shifts (as short time delays) that collectively create a gradient pattern of phase shifts across the {array}, {thus realizing} beamforming. Such short delays can be realized by designing the shape of the elements and can be further tuned by integrating active elements (e.g, varactors) or materials (e.g, liquid crystal (LC)).
The reconfigurability of these short delays leads to phase shifts that add controllability to EM wave propagation that enables {improved} communication and/or sensing performance \cite{annieSIM}, \cite{SIMbeam}. 
 Note that in Fig. \ref{Delayfig}, if the time delays $d_{i,j}$ are all zero, we get a conventional SIS which only introduces short time delays (i.e., phase shifts). Such an architecture has already been investigated in the literature { (cf. \cite{liu2024stackedintelligentmetasurfaceswireless})}.

\subsubsection{Long time delay} 
\label{seclta} A long delay can be applied to EM waves to mitigate propagation effects (such as beam squinting \cite{timedelaybeamsq}) or to optical waves for sensing or network switching. Typically, a time delay is considered large when it is in the order of a symbol duration {(the symbol duration, for a bandwidth of $B$ can be in the order of $1/B$ seconds)}. For instance, a nanosecond delay was integrated in an RIS-assisted wireless system to enhance communication rate in \cite{timedelaybeamsq}. While an element with a short time delay is widely investigated and tested in the literature, a long-delay element is less explored. Fig. \ref{Delayfig} illustrates the conceptual architecture of a DA-SIS, which introduces long time delay values $d_{i,j}$, that are configurable and can be nonzero. Two questions arise related to this conceptualization: Is this feasible? And is it useful? We explore these questions next. We start by detailing two potential approaches that can be adopted to realize an SIS with a large time delay, i.e., a DA-SIS.

\vspace{-0.1in}
\subsection{Feasibility of DA-SIS}
Let us consider an EM wave impinging on a reconfigurable surface. A large time delay can be introduced to the wave using two possible methods: By slowing down its propagation across the medium (passive approach) or by incorporating active circuitry (active approach). We classify the first case as passive since the delay is embedded inherently within the physical medium, and no additional circuitry is required. The second case is considered active, as it involves the usage of active elements such as switchable delay lines or radio frequency (RF) chips that convert the impinging wave into an electrical signal, introduce a programmable time delay, and then retransmit the signal using a secondary emitter. In what follows, we describe the two methods.

\subsubsection{Passive Delay}
\begin{figure}[!t]
\centering
\subfloat[Element-wise delay, where the delay is achieved by changing the applied electrical field on elements' material to realize different propagation speeds.]{\includegraphics[width=0.5\textwidth]{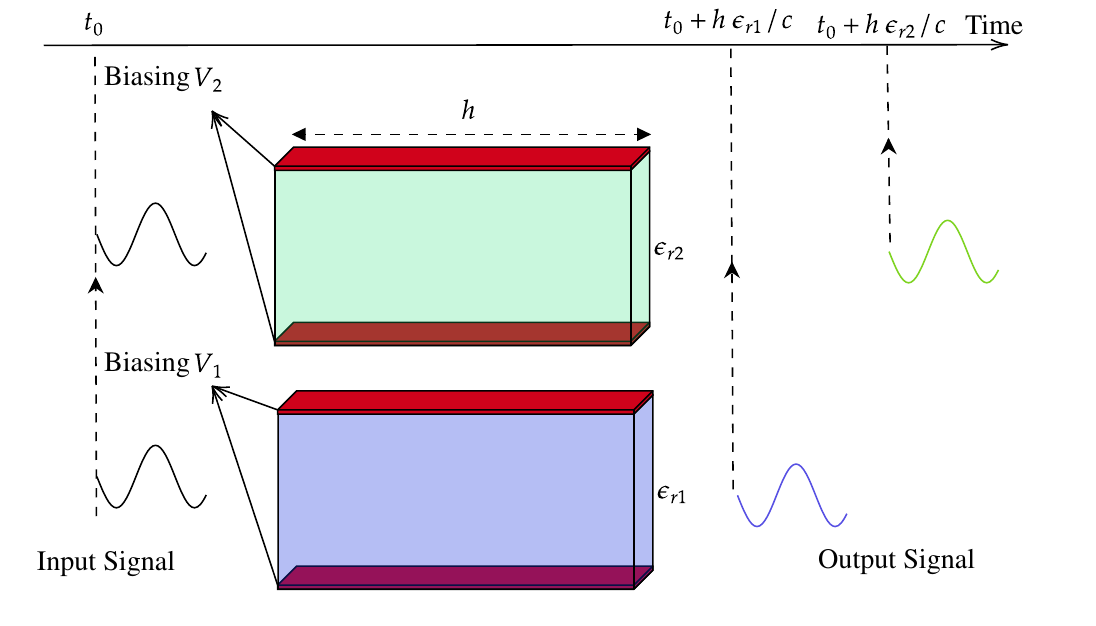}}
\vspace{0.015in}
\subfloat[SIS-wise delay, where materials with fixed characteristics are implemented, and the delays are realized by controlling the wave propagation through the SIS ]{\includegraphics[width=0.5\textwidth]{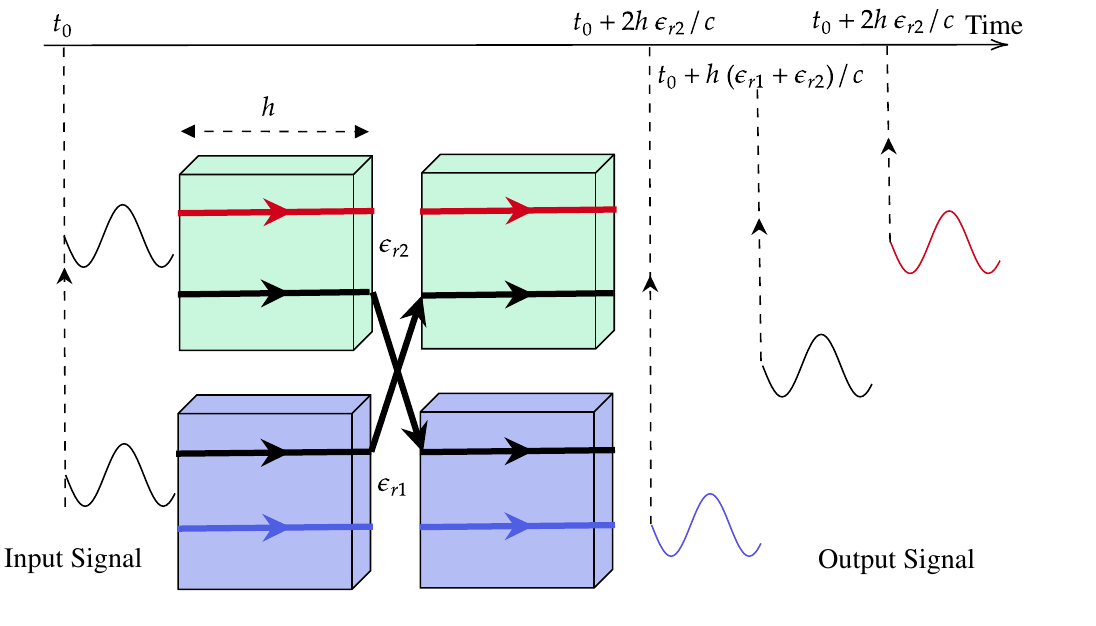}}
\caption{An example of two passive DA-SIS structures: a time delay applied to individual elements (a), and a reconfigurable delay through the whole SIS (b).}
\label{Unitcellarray}
 \vspace{-0.2cm}
\end{figure}

The delay in this case is typically achieved by tailoring the effective propagation velocity through engineered materials or structural configurations that increase the wave's travel {time} or modify its {group} velocity. One common technique in optics relies on photonic crystals that exhibit slow-light behaviour, effectively introducing delay to the optical wave. At lower frequencies, tunable dielectric materials such as LC or ferroelectrics can be embedded within the element structure to achieve controllable time delays up to microseconds \cite{time_delay_ref_Ferroelectric}. By applying a spatially varying DC electric field, the permittivity of these materials can be modulated across the surface, thereby introducing different local delays.
The delay introduced by a material {(relative to free space)} with a permittivity of $\epsilon_r$ and a thickness of $h$ equals to $ h \sqrt{\epsilon_r} / c$, where $c$ is the speed of light in a vacuum. The authors in \cite{time_delay_ref_Ferroelectric} show that ferroelectric materials can achieve permittivity values in the range of $\epsilon_r  \in [300,400]$, which can be used in a multi-layer structure to realize different delays. Fig. \ref{Unitcellarray} (a) shows a sketch of a passive delay element with a controllable delay through changing the $\epsilon_r$ of the used material.

Static delay unit cells can be used to simplify SIS fabrication. In such a design, each unit cell provides a constant time delay, and the delay of the output signals depends on the delay of the constituent units. This is an SIS level delay reconfigurability (instead of the element level), where the output delay can be controlled by directing the EM wave through these materials (through beamforming by using phase shift units {in an electrically large cross-sectional areas and layer spacings}). Hence, the SIS will have multiple propagation paths within the SIS, and the time delay depends on the chosen path. Fig. \ref{Unitcellarray} (b) shows an example of using two different types of materials, each with a different delay {in the order of ns}. At the output of the SIS, three signals can be observed depending on the propagation path: (i) a signal with two short delays (bottom), (ii) a signal with a delay that equals the sum of a short and a long delay (middle), and (iii) a signal with two long delays (top). This design can be extended to provide output signals with more delay options.

{\begin{remark}
 The design proposed for  DA-SIS assisted system comes with the feature of real-time tunability, where a dynamic delay is shown
in Fig. \ref{Unitcellarray} (a) and a static delay is shown in Fig. \ref{Unitcellarray}  (b). Both systems provide real-time tunability through different
control mechanisms. The former is through changing the electric field applied onto the material, while the latter is
through changing the phase-shifts of the elements. 
\end{remark}}

\subsubsection{Active Delay}

Implementing large time delays using active components often results in the design of multi-functional unit cells, capable of providing not only programmable delays but also functions such as power amplification and frequency shifting. For instance, the authors in \cite{time_delay_ref_active} show the design of an optical controllable time delay structure using switches to realize a maximum time delay of 12 ns. This approach provides a high degree of adaptability and control, making it suitable for integrated sensing and communication scenarios. However, these advantages come at the cost of increased design complexity, higher power consumption, and fabrication cost, which must be carefully balanced against the system performance requirements.

{In conclusion, we have discussed various approaches for designing a DA-SIS, inspired by both the propagation behavior of electromagnetic waves in different media and by established signal processing techniques. These approaches can be implemented to realize DA-SIS operation across microwave, millimeter-wave, and terahertz regimes, using different materials or components while adhering to the same underlying principle.}

\subsection{Relevance and Capabilities of DA-SIS}

The integration of time-delay functionality into {RIS layer} components has been previously investigated in the context of RISs, as seen in \cite{timedelayjian,timedelaybeamsq}. While these efforts mark important steps toward spatiotemporal control in reconfigurable surface-assisted systems, their scope remains constrained to RIS frameworks and lacks the broader applicability and flexibility offered by the DA-SIS architecture. In \cite{timedelayjian}, the concept of an active delay-adjustable RIS was introduced through the use of varactor diodes, enabling tunable time-delay control. The authors focused on enhancing RIS-assisted OFDM channels by jointly optimizing the phase shifts and delay parameters to align the reflected signal for the strongest multi-path tap. Their study also highlighted a critical trade-off: The inclusion of practical delay elements can lead to power losses, which must be balanced against the expected performance gains.

Meanwhile, in \cite{timedelaybeamsq}, the challenge of beam squint in wideband systems was addressed by partitioning an RIS into sub-surfaces, each equipped with a common time-delay unit. This design enabled frequency-selective behavior across the RIS, thereby facilitating dynamic beam steering across a wide frequency band.

Although these approaches demonstrate the potential of time-delay integration, they are inherently limited by the structural constraints of RIS systems. In contrast, the DA-SIS introduces a fundamentally different paradigm. By embedding tunable delay elements into the  architecture of SIS, the DA-SIS not only expands the degrees of freedom available for signal manipulation but also introduces the ability to perform temporal filtering on time-series data—unlocking new capabilities for passive, programmable, and low-power spatiotemporal processing.

{The envisioned DA-SIS architecture } enables shifting some of the transceiver processes to the SIS, {paving the way for realizing a range of applications in the wave-domain}  such as channel equalization, filtering, channel coding, etc., one of which is explored in the following case study.
\section{Case Study}

Equalization is a critical process in communication systems to alleviate signal distortions caused by multi-path channels. 
 In this context, we investigate the utility of the DA-SIS as a passive analog equalizer to combat ISI.

To illustrate the potential of the DA-SIS, we consider its deployment at the receiver side of a single-user single-input single-output (SISO) communication link as a simple example (bearing in mind that the idea can be extended to MIMO) . While traditional SISs can be integrated at both the transmitter and receiver ends, we focus on a receiver-side configuration wherein the DA-SIS is used to effectively mitigate  ISI without requiring active circuitry or signal amplification.

In the following section, we explain how the DA-SIS can benefit from the time delay feature to  specifically mitigate ISI. 
\subsection{System Description}
We consider a single-antenna access point {located at $(x,y,z)=(0,0,0)$ and a single-antenna receiver located at $(x,y,z)=(0,100,-15)$ meters} aided by an SIS comprised of $L$ layers containing $N$ elements each and a controller to configure the phase-shifts and delays of the DA-SIS.
Similar to \cite{6gsim}, we consider an SIS with fixed dimensions. Each layer is a square with $\sqrt{N}\times\sqrt{N}$ elements in which $N$ is a square number, and the spacing between layers is {$d_l = 0.75\lambda$ where $\lambda$ is the wavelength at the operating frequency which is set as 28~GHz.}. 
 We denote the diagonal matrix of phase-shifts at each layer $l$ as $\boldsymbol{\Theta}_l$,
and the matrix of transmission coefficients between the $(l-1)$-th and $l$-th SIS layers as $\boldsymbol{H}_l$, with its $(n, \bar{n})$-th entry given according to Rayleigh-Sommerfeld diffraction theory \cite{6gsim}. Each element can introduce a zero or one symbol delay (one bit switchable delay)\footnote{{The delays are restricted to be either $0$ or $1$ symbol duration.}}.

The transmitter sends a sequence of bits $\mathbf{X}_{in}\in \{0,1\}^{M}$ using binary phase-shift keying (BPSK) modulation over a specified $C$-tap vector channel $\mathbf{h} \in \mathbb{C}^{N\times C}$, where $C$ is the number of taps and {$\mathbf{h}$ is the channel between the transmitter and the first layer of the DA-SIS which follows Rician fading with factor $\kappa = 15$. Specifically, we choose $\mathbf{h}$ to be the outer product of the spatial Rician channel and the three-tap (i.e. $C=3$) temporal channel given as $[1, -0.9 e^{j\frac{\pi}{6}}, 0.81 e^{j\frac{\pi}{4}}]$.}
The received signal is the convolution of $\mathbf{h}$ with the input which corrupts the bits with ISI from multi-path. In particular, the SIS $L^{th}$ (outermost) layer receives \begin{align}
\tilde{\mathbf{X}}_L=\mathbf{h}\circledast \mathbf{X}_{in}\in \mathbb{C}^{N\times M+C-1}.
\end{align}
The SIS at the receiver aims to equalize the ISI  by deploying specialized delays at the element level with carefully tuned phase configurations. To model this, we  extend the input matrix  to the SIS into $\mathbf{X}_L=[\tilde{\mathbf{X}}_L, \mathbf{0}_{N\times L}]$, by appending a matrix of zeros (with $N$ rows and $L$ columns) which elongates the time axis to account for the total delay incurred by the SIS (which is at most $L$ symbol delays). We denote $J=M+L+C-1$ as the complete duration of one received sequence considering the delays from the channel and the DA-SIS. 
 The output from the $l^{th}$ DA-SIS layer (i.e. input to layer $l-1$)  is 
\begin{align}
\label{outputtf}
   \mathbf{X}_{l-1}=  \boldsymbol{H}_l \boldsymbol{\Theta}_l(\mathbf{M}_l^1\mathbf{X}_{l}\mathbf{D}^1+\mathbf{M}_l^0 \mathbf{X}_{l}\mathbf{D}^0)
\end{align}
where $\mathbf{D}^0$ and $\mathbf{D}^1$  are time-delay matrices with \begin{align}
\mathbf{D}^0=
\begin{bmatrix}
    \mathbf{I}_{J} &\mathbf{0}_{J}\\
\mathbf{0}_{J}^T &  0
\end{bmatrix},\end{align}
where $\mathbf{I}_{J}$ is the $J\times J$ identity matrix,  $\mathbf{0}_{J}$ is an all-zero vector with dimension $J$, and $\mathbf{D}^1$ is its column-wise cyclic shift to the right. 
Furthermore,  $\mathbf{M}_l^0 ,\mathbf{M}_l^1$  denote binary diagonal masking matrices for each layer $l$ constrained by $\mathbf{M}_l^0 +\mathbf{M}_l^1=\mathbf{I}_{N}$. These matrices model the delay selection for each element, where  $\mathbf{M}_l^0$  selects rows of $\mathbf{X}_{l}$ corresponding to elements that will introduce zero delay, and $\mathbf{M}_l^1$ selects rows corresponding to elements that will introduce one symbol delay.  The mathematical relation in \eqref{outputtf} models the DA-SIS's interaction with an input sequence. {Note that the signal at the receiver's antenna is further corrupted by additive white Gaussian noise.}
\subsection{Optimization for BER Minimization}

To minimize the bit error rate (BER) performance of the end-to-end DA-SIS assisted link, we optimize both the phase shifts and delays of the DA-SIS,  via optimizing $\mathbf{M}_l^0,  \mathbf{M}_l^1$, and $\boldsymbol{\Theta}_l$.{ We use a hybrid scheme of a heuristic method to find the optimized delays then use a stochastic  gradient descent to find the optimal phase shift elements \cite{6gsim}.   }
Initially, we perform randomized initialization of  $\mathbf{M}_l^0$, $\mathbf{M}_l^1$, and $\boldsymbol{\Theta}_l$ {  (where the phases are chosen uniformly and independently at random from $[0,2\pi]$)}. With $\mathbf{M}_l^0$ and $\mathbf{M}_l^1$ held constant, we then apply a gradient descent algorithm to iteratively update the $\boldsymbol{\Theta}_l$, optimizing them to reduce the system’s BER. Once convergence for the fixed delay profile is reached, we reinitialize $\mathbf{M}_l^0$ and $\mathbf{M}_l^1$ randomly and repeat this process to identify a delay-phase combination that yields the lowest BER. The final DA-SIS configuration utilizes the phase profile optimized via gradient descent, paired with the delay settings discovered through the random search.

\subsection{Numerical Results}
{The DA-SIS we consider in this numerical example  consists of $L=2$ layers. The aim of the DA-SIS is to equalize the multi-path channel.}
    
\begin{figure}[t]
\centering
\tikzset{every picture/.style={scale=0.8}, every node/.style={scale=1}}
\input{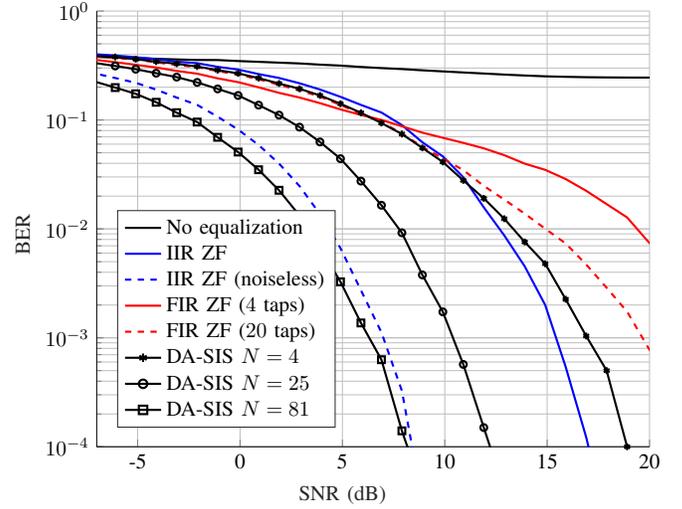}
    \caption{{Numerical BER versus {received} SNR for DA-SIS with different number of elements $N$ and ZF IIR filters {(one with noiseless equalization)} and two FIR approximations (4 taps and 20 taps) as benchmarks. The original ISI corrupted signal BER is also plotted for visualization.}}
    \label{fig:bervnoise}
\end{figure}

    Fig. \ref{fig:bervnoise} shows the performance of the DA-SIS-based analog equalization approach and two conventional digital filtering techniques: An infinite impulse response (IIR) zero-forcing (ZF) filter and two finite impulse response (FIR) approximations of the IIR filter, one with four taps, and the other with twenty taps. {The results highlight and compare the performance of each method. In the case of {DA-SIS} deployment, increasing the number of elements $N$ improves the BER performance. Specifically, for a target BER of $0.001$, the required transmit SNR drops from $62$ dB to $52$ dB as the number of elements increases from $4$ to $81$. This improvement is attributed to two  factors: the enhanced beamforming gain resulting from the larger number of elements, and improved equalization due to more localized temporal solutions across the DA-SIS layers. 
     In the digital filtering deployment, the IIR filter yields the best performance, followed by the approximating FIR filters with $20$ and $4$ taps, respectively. This behavior is expected, as the IIR filter offers an optimal digital solution, and as the FIR filter order increases, its performance approaches that of the IIR filter.}

 {A noticeable performance gap emerges in favor of the DA-SIS with $81$ elements. This superior behavior stems from the ability of the DA-SIS to equalize purely through passive wavefront manipulation; thereby avoiding any amplification of thermal noise. Specifically, the DA-SIS performs equalization in the wave-domain \textit{before} the signal is subjected to additive noise in the receiver electronics.  In contrast, digital equalizers process a signal that has already been corrupted by both ISI and additive noise. As a result, the DA-SIS is able to more effectively suppress ISI without inadvertently amplifying the noise, offering a significant advantage in scenarios where noise sensitivity is a critical constraint.}

 {To further validate this {statement}, we  plot a modified IIR ZF {(noiseless)} benchmark in which the signal is first equalized, then noise is added (noiseless equalization). Compared to the standard IIR ZF approach, this benchmark demonstrates a significant performance gain due to the absence of noise amplification during equalization. Nonetheless, the DA-SIS with 81 elements still outperforms this benchmark.  This superiority stems from the combined effect of DA-SIS’s beamforming gain, resulting in a higher effective SNR, and its wave-domain equalization, which approximates the IIR filter's behavior without incurring noise amplification. In contrast, the IIR filter provides only equalization  without any beamforming gain. {Finally, in the no equalization case, the BER decreases with the increment of the SNR (as the noise is the dominant factor over ISI) until a value after which the BER saturates (called the error floor), due to the dominance of the ISI. }  
}
\vspace{-0.1in}


\subsection{ Discussion: DA-SIS  versus Digital Equalization }

While digital equalization dominates modern systems due to its flexibility and precision, {wave-domain} equalization retains unique advantages in specific scenarios. A major win for wave-domain equalization is time, since signals are adjusted in real time without analog-to-digital conversion (ADC) or digital signal processing (DSP) delays. Furthermore, typical power consumption would be less in wave-domain equalization as it relies either on passive components or simple active circuits without requiring power-hungry high-resolution ADCs/DACs or DSP units. 
While conventional digital equalization techniques such as linear equalizers or decision feedback equalizers can counteract channel distortion at the cost of amplifying received noise or error propagation,  the DA-SIS performs equalization without these limitations as it shapes the  wavefront passively in the wave domain.

\section{Challenges and Opportunities}

As with the incorporation of any new technology, new challenges and opportunities arise. Some of them are listed in this section.
\vspace{-0.1in}
\subsection{Designing and modeling DA-SIS}

Despite the potential benefits of integrating time delay functionality into SIS elements, several practical challenges hinder their implementation. In passive architectures, achieving large and tunable delays often requires materials with high relative permittivity. However, increasing the permittivity is typically accompanied by {a rise in dielectric loss and high mismatch}, which degrades the signal quality and limits the overall efficiency of the surface. This trade-off becomes particularly critical in wideband applications where maintaining low insertion loss is essential. On the other hand, active architectures, while offering greater control and functionality, face limitations in terms of implementation cost and power consumption. These constraints are especially pronounced when scaling to large surfaces that require a dense arrangement of active elements and multilayer structures. The use of amplifiers, delay lines, and digital control circuits adds to the system complexity, which can offset the fundamental advantages of SIS architectures—namely, low power consumption, minimal hardware overhead, and ease of integration. In summary, developing efficient designs for DA-SIS is an interesting research direction to be explored.

\subsection{{Optimization and Abstraction} of DA-SIS}
The DA-SIS has two sets of parameters that need further investigation. One set is the fixed parameters which are the physical SIS properties such as  inter-layer spacing,  inter-element spacing, elements' size, number of elements, and number of layers. Each of these parameters may affect the system performance.
Another set is the dynamic parameters which involve the phase shifts and delay control. Studying the effect of both sets of parameters on performance is important in order to gain insight into the best ways to select (during fabrication) and optimize (during operation) these parameters. {In the case study in this work, we consider fixed SIS physical parameters, while we optimize the phase shift and delay. We optimize the phase shifts  using gradient descent and the delays using randomized search. } While gradient-based optimization serves as a practical and efficient method in our current framework, alternative strategies such as genetic algorithms, reinforcement learning, or joint optimization schemes remain promising future directions.

{Furthermore, to model the DA-SIS in the case study,} we used a recursive formula. There may be simpler models that can subsequently ease the overall optimization of the parameters and unwrap the recursion inherent in the {model in the paper}. Developing a simplified mathematical model of a DA-SIS can thus be another promising direction. {Investigating possible noises that the DA-SIS can add to the signal and modeling these noises is another important direction. } Moreover, the addition of a feedback mechanism in the DA-SIS to perform more elaborate {time-domain} operations is worth looking into.




\subsection{Modes of Operation and Application }


In practical deployments, the delay elements are likely to be discretized, making the study of fractional-delay capabilities particularly relevant. In this context, two modes of operation can be considered: A sub-symbol mode and a multi-symbol mode.  
{In the sub-symbol mode}, the DA-SIS introduces delays that are smaller than the symbol duration, i.e., within a symbol period. This operation is particularly useful in scenarios where (i) the multi-path components in the channel are separated by short delay intervals, requiring fine-grained compensation, and (ii) security-aware or covert communication systems, where introducing fractional delays increases the temporal diversity and complexity of the signal structure—making pattern recognition or signal reconstruction at the eavesdropper side significantly more difficult and time-consuming{\cite{SecuritySubSym}}.

{In the multi-symbol mode}, the delay introduced by the SIS spans one or more symbol durations. This mode is promising for a variety of applications, including (i) delayed symbol alignment in multi-user or cooperative MIMO systems, where intentional time shifts can facilitate interference alignment, user separation, or improved synchronization {\cite{interfereanceallignement}}; and (ii) communication-sensing integration, where delayed replicas of transmitted signals can be used for range-Doppler diversity or retrospective sensing{\cite{CommSens}}.


\subsection{Cost and Complexity}
{Since SIS in general is expected to lower the costs of conventional umMIMO systems, cost analysis and comparisons of such systems is a worthy direction to affirm the motivation behind the SIS technology and different versions of it including the DA-SIS. The overall power budget of the delay elements  (static components see Fig. \ref{Unitcellarray}) along with the phase shifters (ex. varactors) are typically low-cost. 
Secondly, the complexity in terms of design, operation, and optimization may increase as anticipated with the addition of the delays. Furthermore, the DA-SIS is not constricted to be at the receiver, it can be at the transmitter, or at both. This opens up a motley of different applications utilizing the unique ability to adjust spatiotemporal signal characteristics.}
 \subsection{DA-SIS for Wideband Channels}
    
Beam squint remains a challenge for wideband communication systems, arising from frequency-dependent beam steering that degrades beamforming gains. A common solution involves the use of time delays at the transmit antennas to maintain beam alignment across the frequency band. In this context, the DA-SIS design presents a different method for beam squint reduction in wideband channels. By enabling wave-domain signal manipulation, DA-SIS can  replace existing delay-based techniques, offering a flexible solution to maintain consistent beam patterns over wide frequency ranges.

\subsection{Extension to Multi-user settings and MIMO scenarios}
To demonstrate the potential of the DA-SIS, this work focused on a single-user SISO system as a simple proof-of-concept. Looking ahead, {an important} extension lies in adapting the methodology to practical multi-user and massive MIMO (mMIMO) scenarios. In multi-user systems, the proposed DA-SIS framework can aid in interference mitigation {using both temporal and spatial dimensions}. Furthermore, in both mMIMO and umMIMO deployments, the DA-SIS has the potential to reduce the  power consumption {of target applications such as equalization and filtering} {at the transceiver}. This opens promising avenues for enhancing key performance metrics such as energy efficiency, making it a compelling candidate for next generation wireless systems.
\section{Conclusion}
{This paper highlights the relevance of the proposed DA-SIS in equalization. In general, we expect that the time-domain processing capability of the DA-SIS can enable transceiver time-domain processes to be shifted to the SIS, thereby simplifying transceiver design and ensuring scalability for umMIMO and HMIMO systems. } { Future work on design, modeling, optimization, and performance evaluation through theoretical work, simulations, and experiments is required to further understand the potential and limitations of the DA-SIS in umMIMO systems. }
\bibliographystyle{IEEEtran}
\bibliography{IEEEabrv.bib,bib}

\begin{thebibliography}{10}
\providecommand{\url}[1]{#1}
\csname url@samestyle\endcsname
\providecommand{\newblock}{\relax}
\providecommand{\bibinfo}[2]{#2}
\providecommand{\BIBentrySTDinterwordspacing}{\spaceskip=0pt\relax}
\providecommand{\BIBentryALTinterwordstretchfactor}{4}
\providecommand{\BIBentryALTinterwordspacing}{\spaceskip=\fontdimen2\font plus
\BIBentryALTinterwordstretchfactor\fontdimen3\font minus \fontdimen4\font\relax}
\providecommand{\BIBforeignlanguage}[2]{{%
\expandafter\ifx\csname l@#1\endcsname\relax
\typeout{** WARNING: IEEEtran.bst: No hyphenation pattern has been}%
\typeout{** loaded for the language `#1'. Using the pattern for}%
\typeout{** the default language instead.}%
\else
\language=\csname l@#1\endcsname
\fi
#2}}
\providecommand{\BIBdecl}{\relax}
\BIBdecl

\bibitem{emilHMIMO}
Ã.~T. Demir, E.~Björnson, and L.~Sanguinetti, ``Channel modeling and channel estimation for holographic massive {MIMO} with planar arrays,'' \emph{IEEE Wirel. Commun. Lett.}, vol.~11, no.~5, pp. 997--1001, 2022.

\bibitem{6gsim}
J.~An, C.~Xu, D.~W.~K. Ng, G.~C. Alexandropoulos, C.~Huang, C.~Yuen, and L.~Hanzo, ``Stacked intelligent metasurfaces for efficient holographic {MIMO} communications in {{6G}},'' \emph{IEEE J. Sel. Areas Commun.}, vol.~41, no.~8, pp. 2380--2396, 2023.

\bibitem{annieSIM}
Q.-U.-A. Nadeem, J.~An, and A.~Chaaban, ``Hybrid digital-wave domain channel estimator for stacked intelligent metasurface enabled multi-user {MISO} systems,'' in \emph{2024 IEEE WCNC}, 2024, pp. 1--6.

\bibitem{SIMbeam}
N.~U. Hassan, J.~An, M.~Di~Renzo, M.~Debbah, and C.~Yuen, ``Efficient beamforming and radiation pattern control using stacked intelligent metasurfaces,'' \emph{IEEE Open J. Commun. Soc.}, vol.~5, pp. 599--611, 2024.

\bibitem{liu2024stackedintelligentmetasurfaceswireless}
H.~Liu, J.~An, X.~Jia, S.~Lin, X.~Yao, L.~Gan, B.~Clerckx, C.~Yuen, M.~Bennis, and M.~Debbah, ``Stacked intelligent metasurfaces for wireless sensing and communication: Applications and challenges,'' \emph{arXiv 2407.03566}, 2024.

\bibitem{TRaArray}
Y.~Gao, Z.~Wang, H.~Tang, Q.~Luo, and W.~Hu, ``Ultrawideband transmit-reflect-array antenna for {{6G}} communication,'' \emph{IEEE AWP Lett.}, vol.~23, no.~11, pp. 3669--3673, 2024.

\bibitem{timedelaybeamsq}
H.~Sun, S.~Zhang, J.~Ma, and O.~A. Dobre, ``Time-delay unit based beam squint mitigation for {RIS}-aided communications,'' \emph{IEEE Commun. Lett.}, vol.~26, no.~9, pp. 2220--2224, 2022.

\bibitem{time_delay_ref_Ferroelectric}
K.-B. Kim, T.-S. Yun, J.-C. Lee, H.-S. Kim, H.-G. Kim, and I.-D. Kim, ``Integration of coplanar {(Ba,Sr)TiO/sub 3/} microwave phase shifters onto {Si} wafers {TiO/sub 2/} buffer layers,'' \emph{IEEE Trans. Ultrason. Ferroelectr. Freq. Control}, vol.~53, no.~3, pp. 518--524, 2006.

\bibitem{time_delay_ref_active}
R.~L. Moreira, J.~Garcia, W.~Li, J.~Bauters, J.~S. Barton, M.~J.~R. Heck, J.~E. Bowers, and D.~J. Blumenthal, ``Integrated ultra-low-loss 4-bit tunable delay for broadband phased array antenna applications,'' \emph{IEEE Photonics Technol. Lett.}, vol.~25, no.~12, pp. 1165--1168, 2013.

\bibitem{timedelayjian}
J.~An, C.~Xu, D.~W.~K. Ng, C.~Yuen, L.~Gan, and L.~Hanzo, ``Reconfigurable intelligent surface-enhanced {OFDM} communications via delay adjustable metasurface,'' 10 2021.

\bibitem{SecuritySubSym}
R.~Iqbal, M.~Biagi, A.~Zoha, M.~A. Imran, and H.~Abumarshoud, ``Leveraging irs induced time delay for enhanced physical layer security in {VLC} systems,'' \emph{IEEE Wirel. Commun. Lett.}, vol.~13, no.~11, pp. 3147--3151, 2024.

\bibitem{interfereanceallignement}
V.~R. Cadambe and S.~A. Jafar, ``Interference alignment and degrees of freedom of the $k$-user interference channel,'' \emph{IEEE Trans. Inf. Theory}, vol.~54, no.~8, pp. 3425--3441, 2008.

\bibitem{CommSens}
L.~Zheng, M.~Lops, Y.~C. Eldar, and X.~Wang, ``Radar and communication coexistence: An overview: A review of recent methods,'' \emph{IEEE Signal Process. Mag.}, vol.~36, no.~5, pp. 85--99, 2019.

\end{thebibliography}
\end{document}